# Governing Social Media as a Public Utility

Christoph Mueller-Bloch
ESSEC Business School
muellerbloch@essec.edu

Raffaele Ciriello
University of Sydney
raffaele.ciriello@sydney.edu.au

**Abstract:** Social media platforms connect billions, but their business models often amplify societal harm through misinformation, which is linked to polarization, violence, and declining mental health. Current governance frameworks, such as the U.S. Section 230 and the EU Digital Services Act, delegate content moderation to corporations. This creates structural conflicts of interest because misinformation drives engagement, and engagement drives profit. We propose a public utility model for social media governance that prioritizes the public good over commercial incentives. Integrating legislated content removal with democratic content moderation, the model protects free expression while mitigating societal harms. It frames social media as sovereign digital infrastructure governed through democratic oversight, transparent algorithms, and institutional safeguards.

**Keywords:** social media, public utility, governance

# Introduction

While social media connects billions, it also fosters discord, with algorithmic amplification of misinformation linked to polarization [3], violence, and mental health decline.[1] Amid these challenges, a key question is often overlooked: *Who* should govern social media?

As social cohesion erodes, the need for democratic oversight grows more urgent. Corporations like Meta have repeatedly prioritized engagement and ad revenue over the public good, even when aware of the harm their practices cause.[1]

In the United States, governance is shaped by Section 230 and the First Amendment, limiting state intervention and relying on counterspeech to combat falsehoods [8]. The European Union's Digital Services Act imposes more direct obligations for content moderation.[2] Yet both approaches ultimately defer to corporate discretion, perpetuating a conflict of interest: misinformation drives user engagement, and engagement drives profits.

This dynamic constitutes a negative externality – private actions imposing societal costs akin to environmental pollution – which justifies corrective intervention to prevent market failure [9]. The "Bullshit Asymmetry Principle" captures the challenge: refuting misinformation takes significantly more effort than producing it [10].

Treating social media as a public utility curtails profit incentives for misinformation and enables public oversight of content moderation. Like water, electricity, and broadband, social media is critical public infrastructure, serving as a dominant news channel and forum for societal discourse[3]. Building on prior proposals for public alternatives to corporate platforms [6, 7], we develop a model centered on democratic oversight, transparent moderation, legally bounded content removal, and institutional protections for free expression. While ambitious,

---

[1] Horwitz, J. The Facebook Files: A Wall Street Journal Investigation, The Wall Street Journal, https://tinyurl.com/ycytatjd, 2021.
[2] Breton, T. Urgent letter to @elonmusk on #DSA obligations. @ThierryBreton ed., Twitter/X, https://tinyurl.com/mwpynu57, 2023.
[3] Pew Research Center. Social Media and News Fact Sheet, https://www.pewresearch.org/journalism/fact-sheet/social-media-and-news-fact-sheet/, 2024

this proposal seeks to expand what is politically imaginable by reinforcing the democratic foundations of social media governance. Our framework can accommodate private alternatives under public regulation and is adaptable to diverse democratic contexts. Even though the U.S. exemplifies how partisan government intervention offers no viable path towards replacing commercial freedom with civic responsibility, countries such as Germany, Canada, the U.K., and Australia offer proven non-partisan governance models to draw on.

## Adapting the public utility model to social media governance

Governing social media is challenging due to its ever-evolving nature, which defies established regulatory frameworks. While Internet and public broadcasting regulations offer principles – universality, equitable access, quality standards, and resilient funding – these must be adapted to social media's distinct challenges. Chiefly, effective governance requires institutional safeguards to protect free expression and prevent autocratic control.

Social media platforms, whether public or private, face a trade-off between free expression and civil discourse. People value freedom *of* expression but also freedom *from* harmful expression, posing a dilemma when one person's expression harms another. A state monopoly on truth is incompatible with free democracy [9]. So, how can state-governed social media mitigate negative externalities while upholding democratic opinion-forming?

Our proposal distinguishes between content removal and content ranking, with democratically defined and authorized editorial rules. AI can enhance moderation efficiency by detecting illegal content, but auditability and open-source algorithms are essential for trust and accountability.

*Content removal* should be limited to illegal material, enforced by courts. For instance, U.S. First Amendment protections exclude obscenity, child abuse material, incitement to imminent lawless action, true threats, fraud, defamation, and false advertising. Legal misinformation, however, should be addressed through content ranking, not removal, to safeguard free expression.

*Content ranking* should prioritize social connections and user interests over engagement to curb misinformation. Platforms like Reddit[4] use up- and downvotes to rank content, which can curtail misleading yet technically legal "grey-area" content more effectively than simple flagging [1]. Thus far, countries such as the U.S., U.K., Canada, and Germany regulate only illegal content. While this avoids a state monopoly on truth, it fails to address harmful but lawful expression. Our proposed public utility model combines law-driven removal of illegal content with ranking systems that curb the visibility of harmful material without deleting it. This approach limits amplification without censorship or government overreach. Decentralized identifiers[5] can prevent vote manipulation while preserving privacy. Self-sovereign identities[6] enable cross-platform data portability, reducing platform dependence.

Any ranking system entails a trade-off between group preferences and societal cohesion. Conservatives may upvote content that liberals downvote. An algorithm marketplace could enhance in-group satisfaction and reduce inter-group conflict, but risks deepening echo chambers. Aggregated preferences may mitigate fragmentation but marginalize outliers. For public social media, prioritizing consensus through aggregation may be preferable. This approach sidelines, but does not censor, extreme content and leaves it accessible to those who seek it. Integrating social graphs and user interests, such a system supports civil discourse while maintaining an open marketplace of ideas.

Checks and balances are essential to prevent partisan capture by insulating content ranking policies and funding decisions from partisan control. The German public broadcasting model offers a valuable precedent. Germany's constitutional court has repeatedly upheld that public

---

[4] Roose, K. Reddit's I.P.O. Is a Content Moderation Success Story, The New York Times, https://www.nytimes.com/2024/03/21/technology/reddit-ipo-public-content-moderation.html, 2024.
[5] W3C. Decentralized Identifiers (DIDs) v1.0. https://www.w3.org/TR/did-1.0/, 2022
[6] W3C. Identity & the Web, https://www.w3.org/reports/identity-web-impact/, 2025

broadcasting must be pluralistic and independent. Politicians are barred from controlling editorial or funding decisions. Regulatory boards that determine funding and editorial policy are legally required to include members chosen by designated societal groups (e.g., unions and religious organizations), not politicians.

Australia's Online Safety Act 2021 offers a complementary pathway. It empowers an independent *eSafety Commissioner* to mandate the removal of illegal or clearly abusive material across digital services. While it expands regulatory capacity, it does not address "harmful but lawful" expression or algorithmic ranking. Nonetheless, it demonstrates how liberal democracies could build on existing institutions to move toward a public utility model of social media.

Like public broadcasters, public social media should follow content ranking standards overseen by independent institutions to ensure accuracy, impartiality, and educational value. This may require constitutional amendments to create safeguards against partisan interference. Financial independence can be supported through a mix of fees, donations, and revenues from advertising and premium services.

Public and private social media platforms can coexist, much like private TV stations do alongside public broadcasters like the BBC. However, private platforms must be subject to regulatory oversight to ensure responsible ranking and curb engagement-driven harms. Independent regulators – modeled after the UK's Ofcom – could assess compliance using metrics such as *attention allocation* (how algorithms allocate user attention), *revenue attribution* (how attention is monetized), and *impact measurement* (societal consequences of content amplification) [5]. Regulatory tools could range from attention trading mechanisms [9] to more forceful measures like expropriation, antitrust action, or shutdowns in cases of persistent non-compliance.

Another approach – fully compatible with our proposal – would mandate a shared infrastructure supporting multiple ranking systems [8]. Bluesky's *marketplace of algorithms*[7] is one such model, allowing users to choose their preferred ranking systems. A default public-interest algorithm could coexist with commercial alternatives, provided they meet regulatory standards for misinformation control [9]. However, having a multitude of competing social media infrastructures, including a public one, would be preferable, as a unified infrastructure would introduce a single point of failure and make social media as a whole more vulnerable to partisan takeover. Having a public option alongside private social media would be best suited for supporting a pluralistic and democratic society.

**Weighing alternative options**

Alternatives to the public utility model abound. In response to mounting calls for accountability, Meta established an external "Oversight Board" in 2020 to review content decisions and adjudicate user appeals. Likened by CEO Zuckerberg to "a Supreme Court"[8], the board remains structurally and financially dependent on Meta, ultimately leaving policy control in corporate hands. Without accountability beyond Meta itself, the Oversight Board functions more as a PR stunt than a genuine system of checks and balances.[9] By contrast, the public utility model assigns such oversight to *actual* democratic institutions – courts and parliaments – instead of corporate-appointed, pseudo-independent bodies that remain accountable to shareholders rather than citizens.

Recently, Meta shifted from fact-checking to crowdsourced moderation, citing concerns over scalability and the neutrality of fact-checking organizations. However, this shift sidesteps the core issue: Meta retains both the incentive to promote engagement-boosting misinformation and unilateral control over content ranking and removal. Crowdsourcing may broaden participation in moderation, but it does not resolve structural power asymmetries or the absence of public oversight.

---

[7] Graber, K. Algorithmic choice https://bsky.social/about/blog/3-30-2023-algorithmic-choice, 2023
[8] Klein, E., & Zuckerberg, M. Facebook's Hardest Year, and What Comes Next. Vox. Retrieved 29 December 2022 from https://www.vox.com/2018/4/2/17185052/mark-zuckerberg-facebook-interview-fake-news-botscambridge, 2018.
[9] Bentley, J. Meta-funded regulator for AI disinformation on Meta's platform comes under fire: 'You are not any sort of check and balance, you are merely a bit of PR spin', PCGamer, https://www.pcgamer.com/software/ai/meta-funded-regulator-for-ai-disinformation-on-metas-platform-comes-under-fire-you-are-not-any-sort-of-check-and-balance-you-are-merely-a-bit-of-pr-spin/, 2024.

Another proposal, *information fiduciaries*, would impose legal duties of care on platforms [2]. Critics [4] caution that this model requires platforms to balance the conflicting interests of shareholders and society. Because engagement simultaneously drives profit and societal harm, fiduciary platforms may struggle to navigate this conflict of interest, limiting their effectiveness. The public utility model offers a stronger solution by removing profit incentives and aligning operations with societal goals such as equitable access, content integrity, and harm reduction. It also ensures democratic oversight by courts and legislatures, with international cooperation addressing the longstanding regulatory evasion by global corporate platforms.

Recent efforts by President Trump to defund PBS and stack its board with political allies highlight the risk of partisan capture[10], yet they also showcase the relative resilience of democratic institutions: PBS has retained editorial independence, supported by citizen donations and legal protections. In contrast, Meta's content policies rest entirely on executive discretion, requiring no approval beyond Zuckerberg. While imperfect, democratic governance may offer stronger safeguards than corporate self-regulation.

Another proposal envisions a wave of smaller, self-organizing platforms enabled by targeted regulation to address misinformation [11]. While promising, profit-driven smaller platforms funded by advertising would still prioritize engagement, and thus retain incentives to amplify misinformation. A public utility model may address this more effectively by aligning incentives with the public interest.

---

[10] https://theconversation.com/trump-is-aiming-to-silence-public-media-in-the-us-and-if-he-succeeds-his-supporters-here-will-take-note-260584

## The road ahead

Despite its promise, the public utility model faces challenges. More than anything, social media's resilience to capture depends on institutional safeguards rather than on technical design. In robust democracies, public oversight can reinforce accountability and make capture costlier, warranting serious experimentation with the public utility model. Funding may require taxes – a trade-off justified by their contribution to a democratic society. Reconciling national autonomy with global standards remains complex.

Safeguarding free expression and privacy requires robust institutions resilient to partisan or corporate capture. Such capture occurs in many forms: state control, as seen in VKontakte in Russia, WeChat (Weixin) in China, or attempts of the U.S. government to limit enforcement of the European Union's Digital Service Act in Europe, and corporate influence, as evident in the algorithmic moderation and platform bans shaping discourse in Western democracies. These are not opposite extremes but variations of the same problem: concentrated power over online speech. Private platforms have also been weaponized by governments, such as during the Duterte administration in the Philippines, while public ones may be less likely to amplify coordinated misinformation and less vulnerable to interference by foreign governments. Introducing public platforms alongside private ones could diversify the social media landscape, making it less corruptible overall.

Just as today's private platforms are fragmented, so too might public alternatives fragment along jurisdictional lines if they lack interoperability, weakening network effects. Countries will understandably seek to retain editorial control and fiscal sovereignty. Content removal laws vary widely: Holocaust denial is illegal in Germany but protected in the U.S., while abortion service ads are restricted in some U.S. states but permitted in others. A global enforcement framework, akin to SWIFT, could enable cross-border functionality within domestic legal constraints. Without such coordination, divergent regulations may erode cohesion and effectiveness.

In conclusion, while politically contentious, governing social media as a public utility offers a viable path toward a more equitable and resilient digital society. Drawing lessons from robust democracies (such as the U.K., Germany, Canada, and Australia), we propose a model rooted in non-partisan democratic oversight, law-bounded content removal, and transparent moderation. Although the model alone does not guarantee resistance to partisan capture, particularly if constitutional safeguards are lacking, it does resonate with values shared across ideological lines: free expression, checks and balances, rule of law, sovereign infrastructure, and public accountability. Its purpose is not to empower any party or leader, but to strengthen democratic institutions with sovereign digital infrastructure. If social media is to serve the people, it must be accountable to the people.